\documentclass[conference]{IEEEtran}
\IEEEoverridecommandlockouts
\usepackage{cite}
\usepackage{amsmath,amssymb,amsfonts}
\usepackage{algorithmic}
\usepackage{graphicx}
\usepackage{textcomp}
\usepackage{xcolor}
\usepackage{float}  % in the preamble
\usepackage{dblfloatfix}    % To enable figures at the bottom of page
\usepackage{multicol}
\usepackage{threeparttable}

\def\BibTeX{{\rm B\kern-.05em{\sc i\kern-.025em b}\kern-.08em
    T\kern-.1667em\lower.7ex\hbox{E}\kern-.125emX}}
\begin{document}

\title{Linking Opinion Dynamics and Emotional Expression in Online Communities: A Case Study of COVID-19 Vaccination Discourse in Japan*\\
\thanks{This work was supported by JSPS KAKENHI Grant Number 23K28192.}
}

\author{
%\begin{center}
\IEEEauthorblockN{Qianyun Wu}
\IEEEauthorblockA{\textit{School of Computing} \\
\textit{Institute of Science Tokyo}\\
Yokohama, Japan \\
maggie.w.aa@m.titech.ac.jp}
\and
\IEEEauthorblockN{Yukie Sano}
\IEEEauthorblockA{\textit{Institute of Systems} \\
\textit{and Information Engineering} \\
\textit{University of Tsukuba}\\
Ibaraki, Japan \\
sano@sk.tsukuba.ac.jp}
\and
\IEEEauthorblockN{Hideki Takayasu}
\IEEEauthorblockA{\textit{School of Computing} \\
\textit{Institute of Science Tokyo}\\
Yokohama, Japan \\
takayasu.h.aa@m.titech.ac.jp}\\
\and
\IEEEauthorblockN{Misako Takayasu}
\IEEEauthorblockA{\textit{School of Computing} \\
\textit{Institute of Science Tokyo}\\
Yokohama, Japan \\
takayasu@comp.isct.ac.jp}
%\end{center}
}

\maketitle

\begin{abstract}
Social media discourse on COVID-19 vaccination provides a valuable context for studying opinion formation, emotional expression, and social influence during a global crisis. While prior studies have examined emotional strategies within communities and the link between emotions and vaccine hesitancy, few have investigated dynamic emotion changes across collective, community, and individual levels. In this study, we address this gap by conducting an integrated analysis of the evolving collective emotions, community affiliations, and individual emotion changes associated with opinion shifts. Our results show that collective emotions exhibit distinct trends in response to vaccination progress. Emotional compositions differ across communities and respond dynamically to changing pandemic circumstances, potentially reflecting the communities’ influence on users’ opinions. At the individual level, users shifting to pro-vaccine opinions display markedly different emotional changes compared to those shifting toward anti-vaccine opinions. Together, these findings highlight the central role of emotions in shaping users’ vaccination opinions.
\end{abstract}


\section{Introduction}
The discourse surrounding COVID-19 vaccination on social media platforms has attracted significant attention since the onset of the pandemic in 2020. Beyond being a critical real-world issue for promoting vaccination uptake and protecting public health, it also provides a rare, large-scale natural experiment for understanding how humans respond to a global crisis, scientific uncertainty, and rapidly evolving information environments. This discourse is shaped by complicated factors, such as social norms \cite{b1,b2}, political leaning \cite{b3,b4,b5}, social platform algorithms \cite{b4,b5}, and the proliferation of misinformation \cite{b7}, offering a valuable context for examining important multi-disciplinary topics such as opinion polarization, social segregation (echo chambers), and information diffusion.

Understanding the stances toward vaccination (pro-, anti-, and neutral) and how these stances evolve on social media is crucial for promoting informed public discourse and effective health communication. However, this is a complex problem, as vaccination-related discussions are highly dynamic and shaped by constantly changing pandemic conditions and the involvement of diverse actors—such as governments, media organizations, and online influencers. 

In our previous work on COVID-19 vaccination discourse on social media \cite{b5}, we examined the relationship between social communities and individual opinions toward vaccination. We found that users affiliated with pro-vaccine communities were more likely to adopt or maintain pro-vaccine opinions, whereas users in anti-vaccine communities tended to strongly retain anti-vaccine opinions. In the present study, we extend this analysis to the emotional dimension, aiming to understand how communities influence users by shaping their emotions, and how shifts in users’ opinions are associated with corresponding changes in emotional states.

Previous research on the emotional responses to COVID-19 vaccination has primarily focused on two aspects. The first examines how communities shape emotions to influence users \cite{b4,b8,b9}. These studies found that pro-vaccine communities tend to convey positive sentiments, whereas anti-vaccine communities often rely on negative strategies that are perceived as toxic \cite{b4}. The second aspect investigates the association between emotions and vaccination uptake, particularly the roles of trust and the underlying causes of fear and anxiety \cite{b10,b11}. For instance, individuals may feel anxious about the safety and side effects of vaccination \cite{b10,b11}, or exhibit low levels of trust in authorities \cite{b10}.

However, these studies present two main gaps. First, there is a lack of research examining the dynamic associations between emotions, communities, and opinions at a higher temporal resolution. Second, most relevant studies focus on collective-level behavior, while overlooking individual heterogeneity, even though individuals exhibit substantial differences in vaccination preferences and confirmation biases \cite{b2}.

To address these gaps, we analyze a large-scale dataset of tweets and retweets containing the keyword `vaccine' posted in Japanese on X (formerly Twitter) from July 2020 to June 2022, covering the pre-vaccination period through the completion of the booster dose. Our research is conducted at a comprehensive level, covering the macro-level time series analysis to gain knowledge on the dynamic circumstances of vaccination, the mesoscopic-level study to explore the emotional characteristics for different communities, and the micro-level study on the possible associations between a user's opinion and emotion changes. Using retweets, we identify communities and track individual users' dynamic community affiliations within 3-month windows. Using original tweets, we extract the opinion and emotions at both the collective level and the individual level. For this study, we adopt the emotion dimensions defined by POMS (Profile of Mood States \cite{b12}), which include six key emotions: Anger, Confusion, Depression, Fatigue, Vigor, and Tension.

The remainder of this paper is organized as follows. First, we provide an overview of the dynamic changes in opinions, communities, and emotions throughout the vaccination period. Next, we examine the community level, comparing the temporal evolution of emotional compositions across different communities. Finally, we focus on users who changed their opinions—from anti-vaccine or neutral to pro-vaccine, and vice versa—by clustering patterns of emotional changes before and after the opinion shift.

\section{Results}

\subsection{Understanding the overall trend of opinion, emotion and communities.}

\begin{figure*}[htp]
  \centering
  \includegraphics[width=1.0\textwidth]{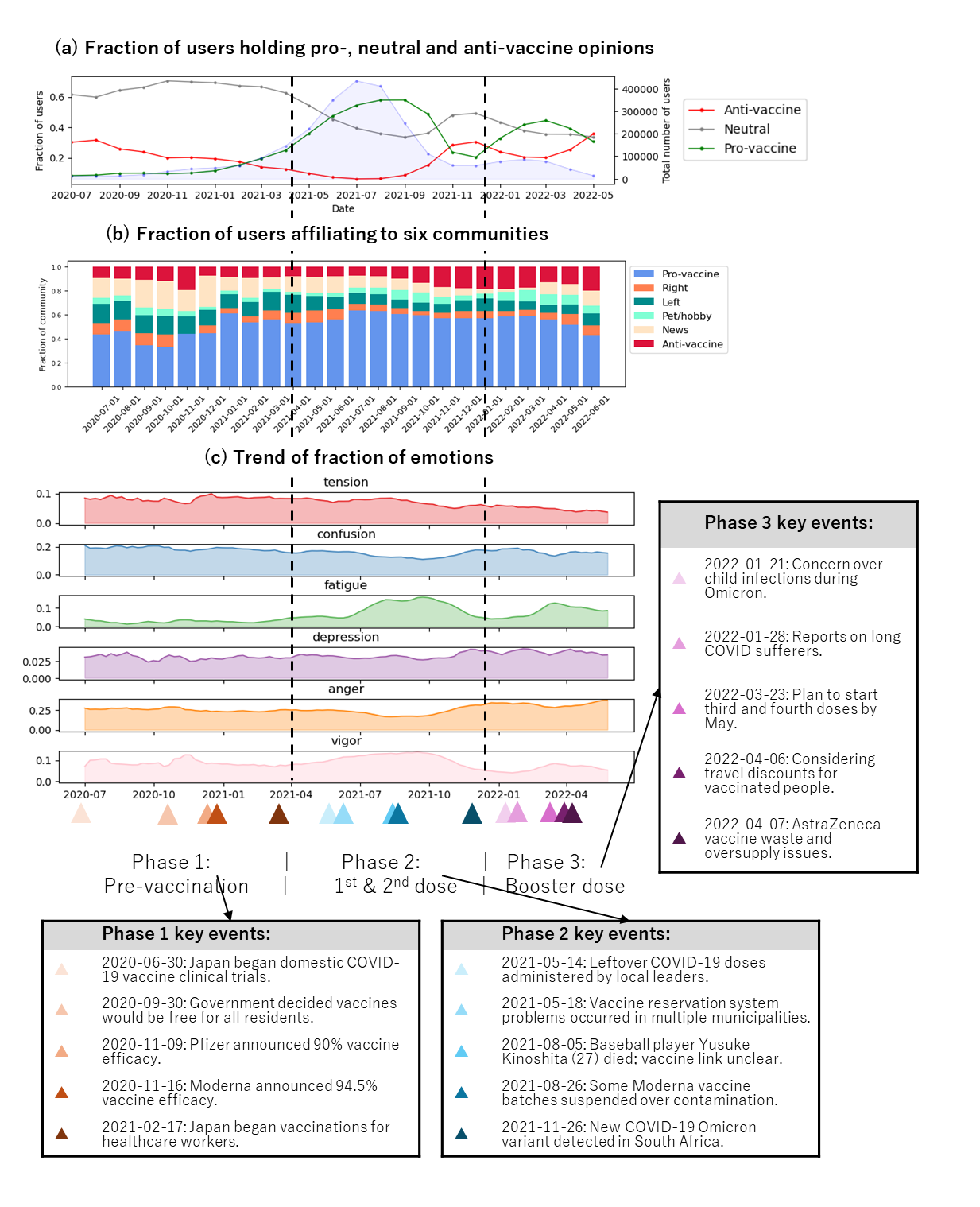}
  \caption{The temporal evolution of collective opinion, social communities, emotions and the hot topics in three vaccination phases. (a) Fraction of users holding pro-, neutral and anti-vaccine opinions \cite{b5}. The y-axis on the left represents the fraction of users with different opinions, while the y-axis on the right represents the total number of users who are active in each time window. (b) Fraction of users affiliating to six communities. \cite{b5} (c) Trend of the fraction of emotions. The black dotted line highlights the boundaries between two vaccination phases. }
  \label{hi}
\end{figure*}

To provide an overview of the evolution of collective opinion, emotion, and community structures over the 2 years following the COVID-19 outbreak, we present \textbf{Fig.~1}, which illustrates the temporal dynamics of opinions, communities, and emotions, along with key topics that may have influenced these changes. We divide the study period into three vaccination phases: \textit{phase~1} — pre-vaccination (July~2020 to March~2021), \textit{phase~2} — first and second doses (April~2021 to November~2021), and \textit{phase~3} — booster dose (December~2021 to June~2022). The \textit{Methods} section details the procedures for detecting individual opinions (pro-, neutral, and anti-vaccination), classifying emotions expressed in tweets into six categories (Anger, Confusion, Depression, Fatigue, Tension, and Vigor) or Neutral, and identifying temporal communities based on the retweet network.

We first extract peaks from the time series and perform keyword analysis to identify key events within each vaccination phase. A peak represents the collective response to a major event, such as news or trending topics. After segmenting the long-term time series into sections, each consisting of one upward followed by one downward trend (see \textit{Methods}–B), we select the most prominent key events and analyze the TF–IDF scores for each word. The TF–IDF metric normalizes a word’s frequency within a specific event by its frequency across the entire period, with higher values indicating keywords that are particularly characteristic of that event. Based on the words with the highest TF–IDF scores, we interpret the corresponding events for the selected peaks and summarize them at the bottom of \textbf{Fig.~1}, where triangles mark the timeline of each key event. The key events in the first phase are mostly related to the effectiveness of vaccination and news about vaccine rollouts, while those in the second phase concern issues surrounding vaccination implementation and the evolving pandemic situation. Events in the third phase are mainly associated with the post-pandemic context. These key events reflect the dynamic evolution of the pandemic and highlight the need for a temporal analysis of opinions, emotions, and communities related to COVID-19 vaccination.

In \textbf{Fig.~1a}, we examine the temporal evolution of opinions, with the y-axis representing the proportion of users holding pro-vaccine (green), neutral (grey), and anti-vaccine (red) opinions \cite{b5}. The method used to classify emotions from tweets and infer each user’s opinion leaning is described in \textit{Methods}–C. During the pre-vaccination stage, users with neutral opinions were dominant. After the start of vaccination, pro-vaccine opinions increased and surpassed neutral opinions between June and October 2021, coinciding with the vaccination rollout. During this period, the fraction of anti-vaccine opinions also decreased markedly. Following the completion of the second dose, the proportions of users holding neutral and anti-vaccine opinions increased again, even exceeding the fraction of pro-vaccine users.

In \textbf{Fig.~1b}, we examine the temporal evolution of communities \cite{b5}. The temporal community detection techniques are explained in the Methods section - D. In overall, the pro-vaccine community remains dominant througout the whole vaccination phase. We can observe an increase of the fraction of pro-vaccine community since the end of 2020, possibly triggered by the news reporting the effectiveness of Pfizer and Modena vaccinations, during which the news community also became larger. On the other hand, the anti-vaccine community is relatively small in size, and it substantially increased at the end of first dose (around October 2021). The increasing activity could be fueled by vaccination misconduct.

Lastly, we evaluate the temporal trends of emotions by randomly sampling 2.5\% of tweets from each day and processing them using a ChatGPT zero-shot model to assign one emotion per tweet. The resulting time series are smoothed using a 3-month moving average. Interestingly, in \textbf{Fig. 1c}, we show that different emotions exhibit distinct patterns: Tension remains consistently high until several months after the vaccination rollout. Confusion is relatively lower during the first- and second-dose phases but increases during the booster-dose phase. Depression remains relatively stable and low compared to other emotions. Anger consistently rises after the completion of the first dose. The trends of Fatigue and Vigor closely follow the vaccination progress: Fatigue corresponds to side effects such as headache and drowsiness, while Vigor reflects cheering, encouragement, and positive engagement with vaccination.

So far, we have examined opinions, emotions, and communities separately. In the following, we evaluate the associations between communities and emotions, as well as between emotions and opinions.

\subsection{Dynamic trend of emotions compositions by communities.}

\begin{figure}[htp]
  \centering
  \includegraphics[width=0.5\textwidth]{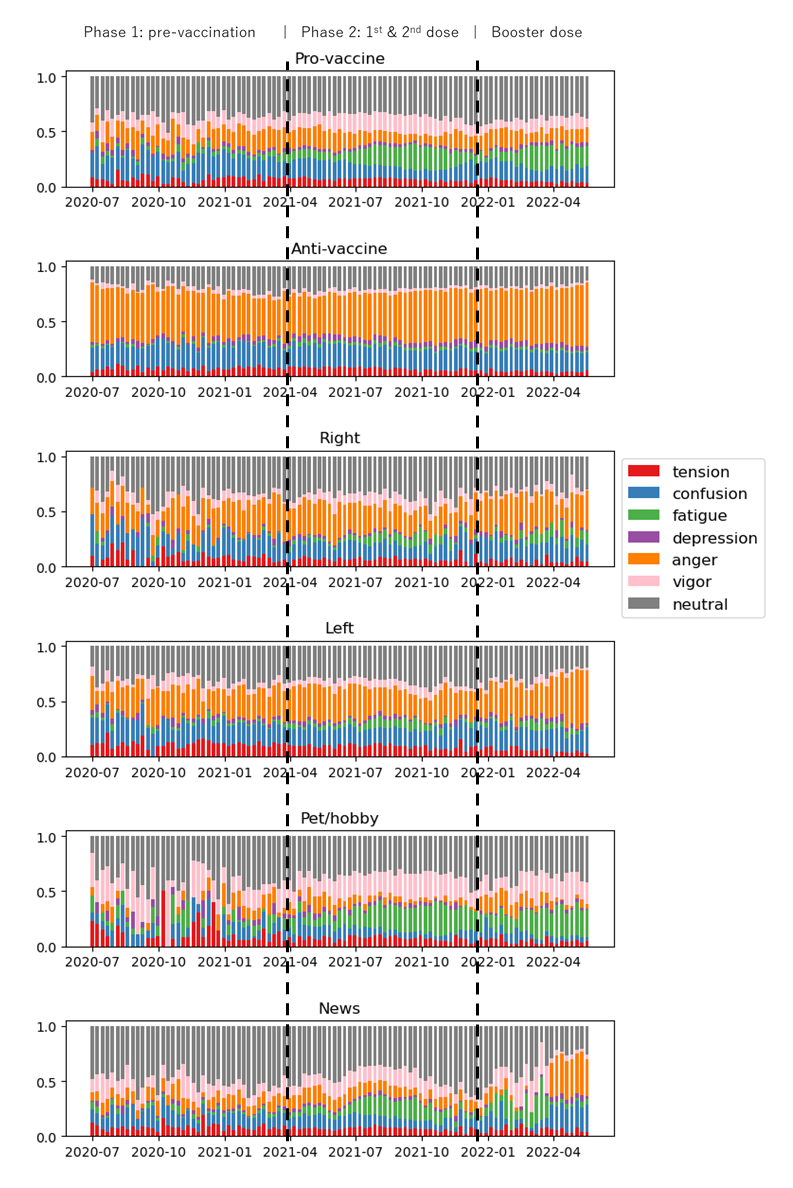}
  \caption{Weekly fraction of emotions by communities. Each sub-figure represents one community. The black dotted lines represent the boundaries between two vaccination phases. }
  \label{hi}
\end{figure}
On top of \textbf{Fig. 1c} which shows the trend of respective emotions, in this sub-section, we use the same 2.5\% randomly sampled data as the analysis for \textbf{Fig. 1c} to present the emotion profiles (fractions of each emotion) within each community. \textbf{Fig.~2} shows an examination of the relative strength of emotions and evaluates whether different communities exhibit distinct patterns of emotional expression. While previous research has investigated sentiment within communities \cite{b4, b8}, our study is novel in analyzing the dynamic trends of emotions based on evolving community structures and showing the nuanced emotion compositions for each community. Our results reveal patterns that not only differ across groups but also change over the course of the vaccination campaign. Moreover, prior work has mainly focused on pro- and anti-vaccine communities, whereas our analysis also includes other communities that may play important roles in vaccination-related discussions.

We first compare the emotional compositions of the pro-vaccine and anti-vaccine communities. Overall, the pro-vaccine community exhibits higher proportions of Neutral, Vigor, and Fatigue, and lower proportions of Anger and Confusion compared to the anti-vaccine community. This pattern suggests that pro-vaccine users generally express more positive attitudes toward vaccination (Vigor), acknowledge fatigue possibly related to post-vaccination experiences, and display lower emotional reactivity. In contrast, the anti-vaccine community appears to be more emotional, manipulating more on toxic narratives emphasizing the risks and uncertainties of vaccination, reinforcing users’ hesitancy.

Examining temporal trends, the pro-vaccine community shows more mixed and volatile emotional patterns, likely reflecting responses to vaccination progress—for instance, increased Fatigue and Vigor and reduced Confusion after vaccination began. By contrast, the anti-vaccine community maintains a stable emotional profile dominated by Anger and Confusion, showing little sensitivity to real-world developments.

Turning to political communities, we find that left-wing and right-wing groups share broadly similar emotional compositions, except that the left-wing community shows slightly higher Tension—possibly reflecting concern over government-led vaccination campaigns—and greater Anger, especially after the booster rollout in 2022. This aligns with our previous findings \cite{b5} that, in Japan, both political camps interact with pro- and anti-vaccine communities, exhibiting a more heterogeneous mix of vaccination opinions than observed in the U.S., where anti-vaccine views are more closely aligned with conservative groups \cite{b4}.

Finally, the remaining two communities—pet and news—display emotional profiles resembling those of the pro-vaccine community, characterized by relatively high Vigor and Fatigue and low Anger during the first and second vaccination phases. However, the news community’s emotions shifted sharply after April 2022, becoming dominated by Anger and Confusion in response to issues surrounding negative news on vaccine or the government such as the AstraZeneca vaccine waste.

\subsection{User's emotion changes accompanying opinion changes.}

Previously, we found that users’ community affiliations were associated with opinion changes toward vaccination \cite{b5}. But what about emotions? Do users exhibit distinct patterns of emotional change when shifting to pro-vaccine or anti-vaccine opinions (see the Methods section for details on opinion detection)?

To address this, we first identify users who changed their opinion according to two criteria:

- The absolute difference in opinion score (see Methods for the definition) between consecutive 3-month time windows is $\geq$ 0.5, to focus on users with substantial opinion changes.

- Users must have posted at least 5 tweets in both adjacent time windows, ensuring sufficient data to calculate each user’s emotion fractions.

We randomly sample up to 100 users from each time window and opinion-change category (e.g., 100 users whose opinion changed from pro-vaccine to anti-vaccine in first to second quarters) to maintain balanced sample sizes across conditions.

Next, we apply K-means clustering to group users based on their emotional composition before and after the opinion change (see a demonstration in \textbf{Fig. 3a}). Each user is represented by a 14-dimensional vector: the first 7 elements correspond to the fractions of the seven emotions in the 3 months preceding the opinion change, and the latter 7 elements correspond to the fractions in the 3 months following the change.

\textbf{Fig. 3b} and \textbf{3c} present the clustering results. Using the elbow method, we select 12 clusters for users shifting to pro-vaccine opinions and 7 clusters for those shifting to anti-vaccine opinions. The bottom-right sub-figure of \textbf{Fig. 3b} and \textbf{Fig. 3c} show the inertia curves used to determine the optimal number of clusters \textit{k}. The black line denotes the inertia values, while the grey dotted line indicates the percentage change in inertia between \textit{k} and (\textit{k}-1) clusters. The red dot marks the local minimum of this inertia difference, corresponding to a noticeable drop in the inertia curve. We select the number of clusters \textit{k} immediately before this drop (red dotted line). Accordingly, we choose \textit{k}=12 for \textbf{Fig. 3b} and \textit{k}=7 for \textbf{Fig. 3c}. In \textbf{Fig. 3c}, multiple local minima appear. To ensure a sufficient number of clusters for the two-sample t-test (see the paragraph after next), we adopt \textit{k}=7.

The sub-figures illustrate that users with similar opinion changes can exhibit distinct patterns of emotional change. For example, among users shifting to pro-vaccine (\textbf{Fig. 3b}), some clusters show an increase in Vigor (center-left and center-right sub-figures) and/or a reduction in Confusion (upper-right and center-right sub-figures). In contrast, for users shifting to anti-vaccine (\textbf{Fig. 3c}), some clusters show an increase in Anger (upper-right and central sub-figures) and/or an increase in Confusion (upper-right sub-figure). These patterns suggest that users’ emotional responses vary systematically depending on the direction of their opinion change.

\begin{figure}[htp]
  \centering
  \includegraphics[width=0.5\textwidth]{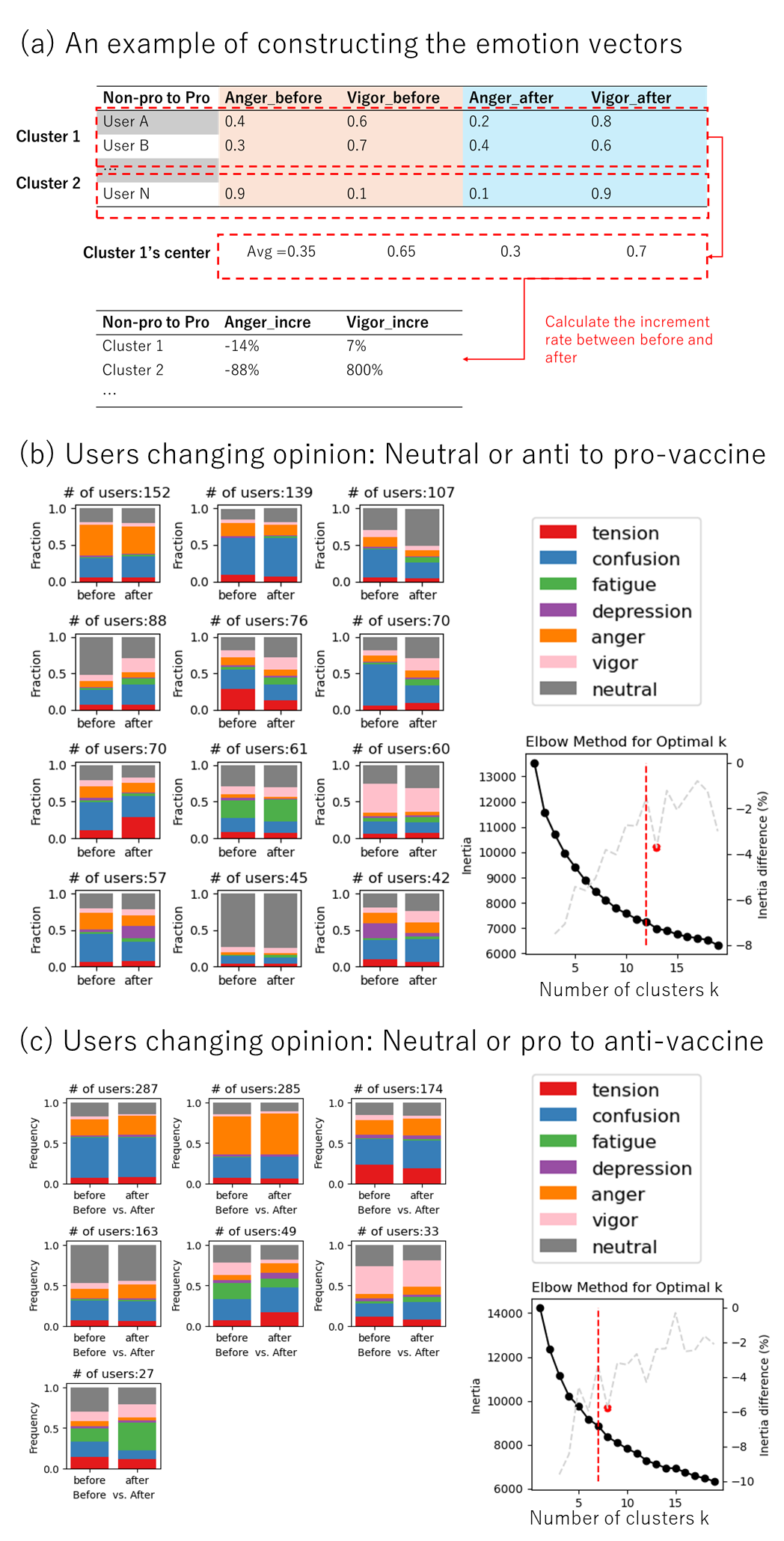}
  \caption{K-means clustering of users who changed opinions. (a) An example of constructing emotion vectors and calculating the increment of emotion for each cluster. (b) Emotion-based clusters for users shifting from neutral or anti-vaccine to pro-vaccine. (b) Emotion-based clusters for users shifting from neutral or pro-vaccine to anti-vaccine. Bar charts show emotion fractions before and after the opinion change; line charts indicate K-means inertia (black curve as represented by the first y-axis) and difference of inertia (grey curve as represented by the second y-axis) used to determine the optimal cluster number via the elbow method. In the line chart, the red dot represents the local minimum point (the turning point), and the red dashed line immediately before the red dot represents the optimal number of clusters that we adopt.}
  \label{hi}
\end{figure}

To statistically assess whether emotion changes differ between users shifting to pro-vaccine versus anti-vaccine opinions, we conduct a two-sample t-test using the increment rate vectors (see \textbf{Fig. 3a}). Specifically, we first calculate the average fraction of each emotion before and after the opinion change for all users within a given cluster. Based on these average emotion vectors for each cluster, we then compute the increment rate of each emotion between the preceding and successive 3-month periods. The resulting increment rates are used to perform the two-sample t-test, comparing emotion changes across clusters of users who changed toward pro-vaccine versus anti-vaccine positions (12 clusters for pro-vaccine, \textbf{Fig. 3b}; 7 clusters for anti-vaccine, \textbf{Fig. 3c}).

\textbf{Table 1} presents the results of the two-sample t-test. Users shifting to anti-vaccine opinions exhibit a significantly greater increase in Anger compared to those shifting to pro-vaccine opinions. Confusion also shows an increase in this group, although the difference is not statistically significant. In contrast, Fatigue, Vigor, and Neutral display significantly larger reductions among users changing to anti-vaccine than among those changing to pro-vaccine.

\begin{table}[htbp]
\centering
\caption{Two-sample t-test of emotion change ratios for users shifting to pro-vaccine vs. anti-vaccine opinions.}
\begin{tabular}{lccc}
\hline
\textbf{Emotion} & \textbf{t (df)} & \textbf{p-value} & \textbf{Cohen’s d} \\ 
\hline
Anger       & $t(17)=3.43$  & .003 **  & 1.75 \\
Confusion   & $t(17)=1.44$  & .167     & 0.72 \\
Depression  & $t(17)=0.01$  & .990     & 0.00 \\
Fatigue     & $t(17)=-2.18$ & .043 *   & -1.09 \\
Vigor       & $t(17)=-2.28$ & .035 *   & -1.13 \\
Tension     & $t(17)=0.02$  & .982     & 0.01 \\
Neutral     & $t(17)=-2.55$ & .021 *   & -1.27 \\
\hline
\end{tabular}
\begin{tablenotes}
\small
\item * $p < .05$, ** $p < .01$, *** $p < .001$
\end{tablenotes}
\label{tab:ttest_emotions}
\end{table}

\section*{Conclusion and Future Work}
 
In this paper, we study dynamic emotion changes across multiple levels: macroscopic (collective time series), mesoscopic (community-level), and microscopic (individual-level). To the best of our knowledge, this is the first study to integrate all three levels to provide a comprehensive understanding of emotion dynamics during the COVID-19 vaccination period.

At the macroscopic level, we observe that the public emotion towards vaccination is highly dynamic, likely influenced by changing vaccination progress and news. We also identify long-term trends in emotions, for example, Anger steadily increases after the first dose of vaccination which may be associated with negative news or controversies surrounding vaccine rollout, Tension steadily decreases since the rollout of vaccination.

At the community level, emotional compositions differ notably across groups and over time. The anti-vaccine community exhibits a stable profile dominated by Confusion and Anger, suggesting that toxic and hostile narratives help reinforce vaccine hesitancy. This aligns with our previous finding \cite{b5} that users in anti-vaccine communities are highly likely to maintain anti-vaccine opinions, potentially due to confirmation reinforcement via emotional content. In contrast, the pro-vaccine community displays a more dynamic and less emotionally intense profile, reflecting varied responses to vaccination progress and related news.

At the microscopic (user) level, we observe significant differences in emotion changes accompanying opinion shifts. Users changing to anti-vaccine opinions are characterized by a notable increase in Anger and significant reductions in Fatigue, Vigor, and Neutral. These findings suggest that emotions are closely linked to opinion formation and may play a key role in shaping individuals’ vaccination attitudes.

These results provide several implications for policymakers and public health professionals seeking to leverage social media for vaccination promotion. First, systematically monitoring emotional responses to vaccination, together with the key topics associated with those emotions, is essential for identifying and addressing emerging public concerns in a timely manner. Second, our results highlight that not only the informational content but also the emotional tone of messages is critical for shaping people’s opinions about vaccination. Finally, when engaging with anti-vaccine communities, it may be effective not only to penetrate these echo chambers to diversify information exposure, but also to directly mitigate anger and confusion through transparent communication, empathetic engagement, and clear explanation of vaccine-related uncertainties.

This study has several limitations. First, it focuses exclusively on Japanese tweets on X (formerly Twitter), which may limit generalizability to other languages, countries, or social media platforms. Second, emotion and opinion inference rely on automated classification, which may not fully capture nuanced or context-dependent expressions. Although the opinion classifier (78\% accuracy) and emotion classifier (81\% accuracy) perform reasonably well, these levels still imply non-negligible classification errors, which could introduce noise into the analysis. Third, we analyze users with a minimum activity threshold and randomly sample up to 100 users per category, potentially biasing results toward more active users. Additionally, the use of 3-month time windows may smooth over rapid changes in opinions or emotions, and retweet-based community detection may not capture all social interactions. Finally, as an observational study, the findings are correlational, and causal relationships between emotions, community affiliation, and opinion change cannot be established. Future work will continue to address these limitations and explore the above directions to gain a deeper and more generalizable understanding of the social media user's opinion and emotion dynamics.

\section{Methods}

\subsection{Dataset}

We collected historical tweets and retweets related to COVID-19 vaccination in Japan using the Academic Twitter API (v2; /2/tweets/search/all/). Tweets containing the Japanese keyword “vaccine” were retrieved, as “COVID-19” is often omitted in everyday discussions. To improve data relevance, we excluded tweets likely to be spam or referring to other vaccines such as HPV, but retained those mentioning influenza, which is frequently discussed alongside COVID-19. The dataset spans from January 2020 to May 2022, covering the pre-vaccination, first, second, and booster dose phases. In total, we collected around 40 million original tweets and 80 million retweets. In the Results section, we will provide more details about the fraction of data used for each analysis.

\subsection{Peak detection}

To detect key events based on peaks in tweeting activity, we construct a time series by aggregating the number of tweets every 10 minutes. As part of data preprocessing, we remove the 10-minute seasonality within a day. Let $S_t$ denote the seasonality at a 10-minute interval, which can be calculated as follows:

\begin{equation}
S_t = \frac{\text{average count of tweets at time } t \text{ (in 10 minutes)}}
           {\langle \text{average count of tweets for each 10 minutes} \rangle}\label{eq}
\end{equation}

Based on seasonality variable $S_t$, we normalize the count time series ${X_t}$ by Equation 2:

\begin{equation}
x'(t) = \frac{x(t)}{S(t)}\label{eq}
\end{equation}

To extract top peaks from long-term time series,  we apply the epislon-tau approach developed by Arthur et al.\cite{b13}. For an upward trend's case, there are two tolerance levels - $\epsilon$ that controls the amplitude of a drop beyond which the trend will be changed to a downward trend, and $\tau$ that controls the time length beyond which the current trend will be ended. \textbf{Fig. 4b} shows an example of detecting the up- and down-trends from the time series. The top point of a up- or down-trend will be regarded as a peak.

\begin{figure}[htp]
  \centering
  \includegraphics[width=0.5\textwidth]{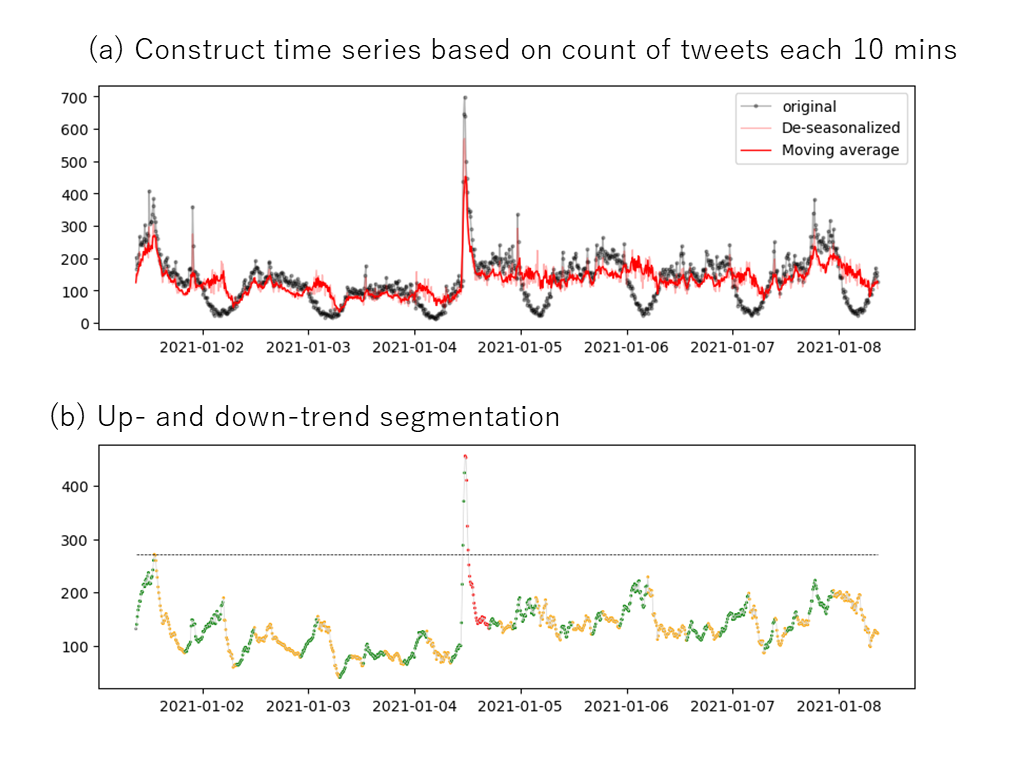}
  \caption{Detecting peaks from time series. (a) Remove seasonality at a 10-minute interval. The black line shows the original time series which shows strong seasonality with lower activity in midnights. The red line represents the de-seasonalized time series. (b) Up- and down-trend segmentation. The green colors represent upward trends, orange represent downward trends but with weaker magnitude, and red represent downward trends from a strong magnitude. }
  \label{hi}
\end{figure}

\subsection{Opinion detection}

In our previous study \cite{b5}, we developed an opinion classification model based on supervised machine learning. We manually labeled 10,000 tweets as pro-, neutral-, or anti-vaccine, and fine-tuned a Support Vector Machine (SVM) classifier that achieved a precision of 78\%. The model utilizes three types of features: pre-trained word embeddings, an opinion-category keyword dictionary, and user community affiliation.

After classifying each tweet into one of the three opinion categories, we inferred each user’s opinion leaning within a specific time window based on the proportion of their tweets in each category (Equation 3) \cite{b5}.

\begin{equation}
O_t^i = \frac{NP_t^i - NA_t^i}{NP_t^i + NN_t^i + NA_t^i},
\end{equation}

where $NP_t^i$, $NN_t^i$, and $NA_t^i$ denote the number of pro-, neutral-, and anti-vaccine tweets, respectively, posted by user \textit{i} during time window \textit{t}.

Here, we define $O_t^i > 0.3$ as pro-vaccine, $O_t^i < -0.3$ as anti-vaccine and $-0.3 \leq O_t^i \leq 0.3$ as neutral.

\subsection{Retweet network and temporal community detection}

In our previous study \cite{b5}, we built retweet network snapshots based on retweets, with nodes representing individual users who retweeted the other, links representing retweeting relationship and link weight representing the frequency of retweets in a specific time window. 

We adopted an Ensemble Louvain method \cite{b18} which is based on the Louvain method, but mitigating the instability caused by random ordering of nodes during grouping. Specifically, the basic Louvain method optimizes modularity, which measures the density of connection within a community as compared to random connections between nodes. The Ensemble Louvain method addresses the randomness issue by applying the Louvain method on the same network for N times. Only nodes that are clustered into the same community for more than n times (i.e., n=90 out of N=100 times) will be regarded as the true community. 

Based on all retweets from the 2-year's time, we built a static network and detected 6 key communities: pro-vaccine, anti-vaccine, left-wing, right-wing, news and pet. The identity of communities is determined based on hashtags, profiles of influencers, and keywords. 

Then we further track the temporal evolution of these six communities. We build temporal network snapshots by aggregating retweet links every 3 months. For each network snapshot $G_t$ at time $t$, we apply the same Ensemble Louvain method \cite{b18} to detect a set of communities \{$C_t^i$\}. To match the communities in adjacent timesteps, we calculate the fraction of overlap between nodes in community $C_{t-1}^i$ and community $C_{t}^j$. Based on the rules defined by Chen et al \cite{b19} and our previous study \cite{b5}, we match communities that have the highest overlap and determine the type of community evolution: birth, death, grow, shrink, merge, split, partial merge/split. For example, community $C_{t}^j$ consists of 50\% of nodes from community $C_{t-1}^a$ and 50\% of nodes from community $C_{t-1}^b$, then we can determine that the previous two communities at time ($t-1$) merge into a bigger community at time $t$. 

Furthermore, to associate these temporal communities with the six key communities detected based on the static network, we select the top 100 influencers from each of the six communities and track their affiliation in temporal communities. For example, the core users of pro-vaccine community are mostly affiliated to community $C_{t-1}^a$ and $C_{t-1}^b$ at time ($t-1$), so that we can label both $C_{t-1}^a$ and $C_{t-1}^b$ as the pro-vaccine community, even though they may correspond to different social groups such as official accounts and medical workers. In Fig. 1b we show the temporal evolution of the six key communities.

\subsection{Emotion classification}

Emotion detection technologies have evolved significantly over the past decade, progressing from dictionary-based approaches \cite{b14,b15}, to supervised and unsupervised machine learning methods using various features \cite{b5}, and more recently to large language models (LLMs) \cite{b16}. LLMs offer a deeper understanding of context, enabling the capture of nuanced and complex emotional expressions, including sarcasm. Previous studies have shown that zero-shot opinion detection and classification using LLMs can achieve reasonable performance \cite{b17}.

In this study, we leverage the ChatGPT 4o-mini model in a zero-shot setting, asking it to output a single emotion (or Neutral) for each text. During sample checks, we observed that, in the context of vaccination, Tension is often misclassified as Confusion, since unknown vaccine outcomes can trigger tension responses. To address this, we explicitly define Confusion and Tension in the query to distinguish them clearly. The query reads:

\textit{"For each following tweet delimited by ';', first identify the emotion most strongly expressed, then output one of [Anger, Depression, Vigor, Fatigue, Confusion, Tension, Neutral]. Distinguish Confusion (uncertainty or lack of understanding) from Tension (unexpected events or concern about outcomes). Only output in the format 'id: emotion' without explanation."}

We set the temperature parameter to 0 to ensure deterministic responses. To validate the classification, we randomly sampled 100 tweets and found an accuracy of 81\%, which we consider reasonable for our analysis.

\end{document}